# Adaptive User Interface Generation Through Reinforcement Learning: A Data-Driven Approach to Personalization and Optimization

Qi Sun
Carnegie Mellon University
Pittsburgh, USA

Yayun Xue
Pratt Institute
New York, USA

Zhijun Song
Parsons School of Design
New York, USA

*Abstract*— This study introduces an adaptive user interface generation technology, emphasizing the role of Human-Computer Interaction (HCI) in optimizing user experience. By focusing on enhancing the interaction between users and intelligent systems, this approach aims to automatically adjust interface layouts and configurations based on user feedback, streamlining the design process. Traditional interface design involves significant manual effort and struggles to meet the evolving personalized needs of users. Our proposed system integrates adaptive interface generation with reinforcement learning and intelligent feedback mechanisms to dynamically adjust the user interface, better accommodating individual usage patterns. In the experiment, the OpenAI CLIP Interactions dataset was utilized to verify the adaptability of the proposed method, using click-through rate (CTR) and user retention rate (RR) as evaluation metrics. The findings highlight the system's ability to deliver flexible and personalized interface solutions, providing a novel and effective approach for user interaction design and ultimately enhancing HCI through continuous learning and adaptation.

## I. INTRODUCTION

In recent years, with the rapid development of artificial intelligence technology, the demand for intelligent user interface (UI) design has gradually increased [1]. The user interface directly affects the user experience, and efficient and intuitive interface design can significantly improve the fluency and satisfaction of user operations. However, traditional interface design requires professional designers to invest a lot of time and energy in manual design and optimization, which makes it difficult to meet the personalized needs of users. Based on this, the adaptive generation technology of user interface combined with reinforcement learning (RL) technology has gradually attracted widespread attention [2]. By introducing reinforcement learning, the system can automatically adjust the interface layout under the real-time feedback of users, gradually meet the user's usage preferences, and provide a new automated and adaptive method for interface design [3].

The core advantage of reinforcement learning is that it continuously optimizes strategies through interaction with the environment [4], which provides a high degree of flexibility for user interface generation. Interface generation based on reinforcement learning can obtain user behavior feedback through intelligent agents, and dynamically adjust on this basis, so as to adapt to the personalized needs of users [5]. For example, RL algorithms can continuously learn the user's behavior patterns and usage habits in the interface, and optimize the layout, button size and other element configurations of the interface in real-time according to the user's feedback. Compared with the traditional interface design mode, this adaptive generation technology can achieve agile response to user needs and provide a highly personalized operation experience [6].

In addition, the user interface generation technology combined with reinforcement learning can greatly improve the design efficiency of the interface. The traditional user interface design process usually requires multiple rounds of user testing and manual debugging, and the design cycle is long. The reinforcement learning system can automatically iterate and generate multiple versions of interface styles in a simulated environment through agents and reward mechanisms [7], so as to quickly find the optimal design solution. Such efficiency advantages not only greatly shorten the cycle of interface design and testing but also reduce design costs, allowing designers to focus on innovation and complex design tasks. The interface generated based on RL technology can be adaptively optimized for the operation preferences of different users while maintaining uniformity and has high application potential.

The application of reinforcement learning in the adaptive generation of user interfaces also shows excellent accuracy and stability. By building a user behavior model and designing a reasonable reward function, the system can accurately capture the user's operation preferences in the interface, avoiding the problem of trial and error in traditional methods. During the training process, the reinforcement learning model can continuously adjust the generation strategy according to user feedback, realize the precise optimization of interface elements, and thus improve the user experience. For example, the system can make real-time adjustments based on information such as user stay time and click frequency so that the generated interface is more in line with the user's usage habits. This data-driven optimization method makes the interface generation more in line with actual needs and has significant practical value.

The advantages of reinforcement learning in the field of user interface generation are not only reflected in personalization and efficiency but also in good scalability. RL algorithms can be combined with other models (such as generative adversarial networks) to achieve multi-level interface generation and optimization [8]. For example, after the basic interface is generated, the system can use generative adversarial networks to further improve the interface quality to better meet the user's aesthetic needs. In addition, the structured design of the RL model is easy to expand to different platforms and application scenarios. For example, in the e-commerce field, an adaptive interface could revolutionize online shopping by personalizing the browsing experience. For instance, frequent shoppers might see trending items or exclusive offers prominently displayed, while budget-conscious users could be presented with discounts and economical bundles. This dynamic adjustment not only improves user satisfaction but also drives higher conversion rates. In education, adaptive interfaces could transform digital learning platforms by customizing content delivery based on user preferences and interaction patterns. A student struggling with a concept might receive additional visual aids or interactive simulations, while another who learns better through text might see more detailed articles and notes. Similarly, in healthcare, telehealth applications could optimize patient engagement by adjusting their dashboards. For example, a user managing chronic conditions could have their interface prioritize medication reminders and health metrics, whereas older adults might benefit from simplified navigation and larger text sizes. These examples underline the versatility and impact of adaptive interfaces across diverse fields.

The proposed system aligns with several key trends in Human-Computer Interaction, further underscoring its relevance and potential. Predictive interfaces, a growing trend in HCI, seek to anticipate user needs by leveraging data and machine learning. The adaptive interface could extend this capability by preloading likely actions or features, such as offering shortcuts to commonly accessed tools or predicting the next step in a workflow. Another trend, zero-interaction design, emphasizes reducing the cognitive load on users by minimizing manual input. The adaptive system supports this by automating repetitive tasks and adjusting the interface layout based on usage patterns. Lastly, continuous learning interfaces, which evolve based on user interaction, are increasingly popular. The integration of reinforcement learning in the proposed system ensures that it adapts not only to individual users but also improves its performance over time, offering a dynamic and personalized user experience. By aligning with these trends, the system showcases its relevance to contemporary HCI advancements and its potential to drive future innovation.

## II. RELATED WORK

The advancement of adaptive user interface generation has been closely tied to recent developments in machine learning, deep learning, and data-driven personalization methodologies. Duan et al. [9] introduced an Interface Generation Tree Algorithm leveraging deep learning to optimize design efficiency and aesthetics, a foundation closely aligned with the objectives of adaptive UI generation. Similarly, Wang et al. [10] applied predictive modeling techniques to dynamically optimize system configurations, which can inform real-time UI adaptations based on user interactions. Feng et al. [11] demonstrated the potential of generative models to enhance data efficiency in machine learning tasks. Their approach provides valuable insights into data-driven personalization strategies, which are crucial for adaptive UI systems. Shen et al. [12] further explored leveraging semi-supervised learning to achieve robust performance under limited data conditions, a challenge often encountered in real-world adaptive interface deployments.

The integration of deep learning architectures such as encoder-transformer models, as proposed by Chen et al. [13], highlights their capability to ensure high-quality adaptive decision-making in UI configurations. Additionally, Liu et al. [14] addressed optimization challenges in feature learning, presenting methodologies that mitigate common issues such as over-smoothing, which can similarly arise in reinforcement learning-driven systems.

Graph Neural Networks (GNNs) have been explored for their potential in structuring and processing complex relationships. Du et al. [15] demonstrated the use of GNNs for intelligent reasoning, offering principles that can enhance adaptive UI mechanisms by structuring interaction data. Wei et al. [16] expanded on this by introducing self-supervised approaches for feature extraction, reinforcing the value of GNNs in data representation for adaptive systems. Liang et al. [17] focused on dimensionality reduction techniques, which support scalable and efficient processing of large interaction datasets, a key requirement for adaptive UI systems. Yan et al. [18] proposed methods for converting complex datasets into interpretable forms, directly contributing to the usability and transparency of adaptive systems. Finally, Xiao [19] highlighted the robustness of self-supervised learning for few-shot classification, an approach that ensures personalization and adaptability in environments where user feedback data is sparse or intermittent.

These studies collectively provide the theoretical and practical underpinnings for the design of adaptive user interface systems, focusing on reinforcement learning, data-driven optimization, and advanced machine learning methodologies to deliver dynamic, user-centric solutions.

## III. METHOD

The method of this study is based on the core idea of reinforcement learning (RL) and realizes the adaptive generation of user interfaces by building intelligent agents [20]. The main goal of the reinforcement learning model is to learn the optimal strategy so that the intelligent agent can continuously optimize the interface layout and element configuration driven by user feedback [21], thereby maximizing the user experience. First, the state of the interface generation system is set to $s$, that is, the state of the current interface layout; action a represents the adjustment operation performed by the system on the interface, such as changing the button size, layout or color; the reward function $r$ is used to measure the user's satisfaction with the current interface. The goal of reinforcement learning is to maximize the cumulative reward $R$, which is defined as follows:

$$R = \sum_{t=0}^{T} \gamma^t r_t$$

$\gamma \in [0,1]$ is the discount factor, which indicates the importance of future rewards. In each round of training, the intelligent agent observes state *s* and selects an action *a* based on strategy $\pi(a|s)$, then obtains reward *r* based on user feedback and enters the next state $s'$. By continuously updating strategy $\pi(a|s)$, the system can gradually learn the optimal interface adjustment strategy.

In the specific implementation, the deep Q network (DQN) is used to estimate the value of each state-action pair, that is, the *Q* value. The *Q* value function $Q(s,a)$ represents the expected reward that can be obtained when taking action *a* in state *s*. In order to optimize the *Q* value function, the Bellman equation is used for iterative updates, and the formula is as follows:

$$Q(s,a) = r + \gamma \max Q(s',a')$$

In each iteration, DQN updates the *Q* value based on the reward information of the current state *s* and action *a*, and maximizes the *Q* value to find the best action sequence that can improve the user experience. To ensure the training stability of the model, the experience replay mechanism is introduced to store the samples of each training in the memory pool, and the temporal correlation of the samples is broken by random sampling, which further improves the generalization ability of the strategy.

In addition, this study ensures that the generated interface meets user preferences by designing a reasonable reward function. The reward function is set to $r = f(u)$, where *u* represents user feedback, such as the user's click frequency, dwell time, and other specific interaction data. The larger the value of the reward function, the closer the generated effect of the current interface is to the user's ideal state. The intelligent agent maximizes the cumulative reward in each decision-making process to achieve the adaptive generation of the interface and finally obtains the best interface configuration that meets user needs [22].

Through continuous optimization of reinforcement learning strategies, the intelligent agent can gradually adjust the details of the interface in multiple rounds of training to meet the personalized needs of users. Through the above methods, this study realizes the generation of user interfaces based on reinforcement learning, providing a flexible and intelligent solution for interaction design.

## IV. EXPERIMENT

### A. Datasets

In order to implement the adaptive generation of user interfaces experiment, this study selected the OpenAI CLIP Interactions dataset. This dataset was jointly developed by OpenAI and multiple research institutions and is mainly used to study the interactive behavior of users in different interfaces. The dataset contains rich information on interface elements, including buttons, text boxes, icons, sliders, etc., and records various user operation data in the interface, such as click location, dwell time, and operation sequence. Each interface screenshot is annotated, covering the detailed attributes of the interface layout and the specific operation trajectory of the user, providing a detailed reference for analyzing user preferences and behavior patterns. In addition, the CLIP Interactions dataset contains interface samples of various application categories, including social media, e-commerce, education platforms, and health applications, which enables the study to analyze the usage habits of users in different application fields. The dataset leverages the Linked Data methodology to unify diverse data formats [23], promoting seamless integration and analysis across domains such as machine learning and artificial intelligence. This strategy eliminates data silos, enhances dataset diversity, and fosters the creation of more precise and scalable AI solutions. Through this dataset, the reinforcement learning model can learn the user's operation rules from it, and then generate an interface layout and interaction method that better meets user needs. With this dataset, this study can verify the effectiveness of adaptive interface generation technology and lay a data foundation for building a more intelligent and personalized user interaction interface.

### B. Experimental Results

In the experimental part, in order to comprehensively evaluate the adaptive generation technology of user interfaces based on reinforcement learning, this study selected five representative comparison models. These models include the interface optimization method based on the Multi-Armed Bandit (MAB) algorithm [24], which gradually optimizes the user experience by exploring different interface layouts; the Bayesian Optimization model, which selects the optimal interface parameter configuration through probability distribution; the interface generation based on the Markov decision process (MDP) [25], which is used to simulate the continuous decision-making of users in the interface; the Policy Gradient algorithm [26], which is used for deep learning-driven interface generation; and the Collaborative Filtering [27], which generates personalized interfaces through the similarity of preferences between users. These models cover a variety of generation ideas from classical methods to deep reinforcement learning models [28], providing a multi-angle comparison for evaluating adaptive generation technology.

In terms of evaluation indicators, this paper selected click-through rate (CTR) and user retention rate (RR) as the main evaluation indicators. The click-through rate is used to measure the attractiveness of the generated interface elements to users, reflecting whether the interface design meets user needs. The higher the click-through rate, the more attractive the interface is. The user retention rate is used to evaluate the user's stay time on the interface and the frequency of revisits, which can directly reflect the user stickiness of the interface. These two indicators can clearly understand the actual effect of different models generating interfaces, and provide quantitative comparison support for the final adaptive generation method.

Table 1 Experimental Results

| Model | CTR | RR |
|---|---|---|
| MAB | 0.65 | 0.72 |
| Bayesian Optimization | 0.68 | 0.74 |
| MDP | 0.70 | 0.76 |
| Policy Gradient | 0.72 | 0.78 |
| Collaborative Filtering | 0.69 | 0.75 |
| Ours | 0.78 | 0.83 |

The reinforcement learning-based model (Ours) achieves a CTR of 0.78 and an RR of 0.83, significantly outperforming other models in user interface adaptive generation tasks, as shown in Table 1. These results highlight its superior ability to attract clicks and retain users compared to methods such as the multi-armed bandit algorithm (MAB), Bayesian optimization, Markov decision process (MDP), and policy gradient algorithm. The policy gradient algorithm and MDP show moderate performance (CTR: 0.72, 0.70; RR: 0.78, 0.76) due to their partial capacity to adapt to user behaviors. However, the policy gradient algorithm is limited by fixed update rules, while MDP relies on static state transitions, reducing flexibility in dynamic environments. MAB and Bayesian optimization perform less effectively (CTR: 0.65, 0.68; RR: 0.72, 0.74) because of their limited ability to capture user preferences or adapt to interaction feedback.

Reinforcement learning's ability to optimize interface designs through iterative training, real-time adjustments, and reward mechanisms makes it highly effective for generating user-centered designs. It dynamically aligns interface features with user preferences, surpassing the limitations of other models.

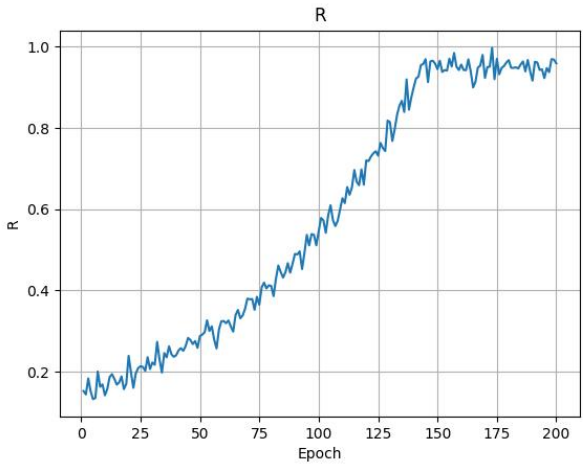

Figure 1 Rising Trend of Activation function

Figure 1 illustrates the upward trend of the activation function during training, demonstrating that as training progresses, the reinforcement learning model converges to a higher policy level. Early in training, the model exhibits significant fluctuations, but with more iterations, the reward values stabilize, reflecting the refinement of the model's strategy for generating user interfaces. This trend highlights the model's ability to optimize the interface layout and configuration dynamically through iterative state-action feedback loops.

The rising reward values confirm the reinforcement learning model's effectiveness in adaptive interface generation and underscore the critical role of the reward mechanism in personalized design. By integrating metrics like CTR and RR, the model effectively captures user preferences, resulting in interfaces that attract clicks and sustain long-term usage. This improvement validates the practical value of reinforcement learning for interface design. Additionally, the model demonstrates robustness and adaptability through repeated training and optimization. The stabilization phase in Figure 1 indicates that the model identifies optimal strategies for interface generation, accommodating user habits and delivering personalized experiences. This capability is pivotal for practical human-computer interaction applications, enabling systems to respond dynamically to evolving user needs. Overall, the study leverages reinforcement learning to achieve efficient and user-centric interface design, significantly enhancing design quality and user satisfaction. With continued advancements in reinforcement learning, such adaptive systems are poised to deliver intelligent and personalized interaction experiences across diverse platforms.

Furthermore, Table 2 presents the impact of different learning rates on experimental outcomes, offering additional insights into the model's optimization process.

Table 2 The impact of different learning rates on experimental results

| Optimizer | CTR | RR |
|---|---|---|
| Lr=0.005 | 0.77 | 0.81 |
| Lr=0.003 | 0.76 | 0.82 |
| Lr=0.002 | 0.77 | 0.81 |
| Lr=0.001 | 0.78 | 0.83 |

Table 2 illustrates the impact of different learning rates (Lr) on the reinforcement learning model's performance, focusing on click-through rate (CTR) and user retention rate (RR). The results reveal that learning rates significantly influence the model's training dynamics and final outcomes, particularly in adaptive user interface generation tasks. At a learning rate of 0.005, the model achieves a CTR of 0.77 and an RR of 0.81, indicating good but suboptimal performance. The higher learning rate accelerates training and feedback assimilation but introduces instability, leading to oscillations and missed fine-tuning opportunities. This limits the model's ability to fully capture nuanced user preferences. With a learning rate of 0.003, CTR is 0.76, and RR improves to 0.82, suggesting better long-term user retention. The reduced step size enhances stability but slightly lowers responsiveness to user click behaviors, highlighting a trade-off between stability and sensitivity.

A learning rate of 0.002 yields balanced results (CTR: 0.77, RR: 0.81), combining moderate responsiveness with improved stability. This rate mitigates oscillations while maintaining effective preference capture, making it suitable for tasks requiring a balance of click preference and retention.

The best performance is observed at a learning rate of 0.001, where CTR reaches 0.78 and RR 0.83. The lower rate enables detailed parameter adjustments, smooth strategy updates, and precise user behavior capture, resulting in enhanced personalization and prolonged user engagement. Although slower, the increased accuracy and stability are crucial for adaptive interface generation tasks. In conclusion, the learning rate plays a pivotal role in determining the model's CTR and RR performance. Lower learning rates improve stability and accuracy, essential for a user-friendly interface experience. These findings underscore the importance of tuning learning rates to balance training speed, stability, and task-specific requirements, ensuring optimal reinforcement learning performance in human-computer interaction applications.

In addition, we also give the impact of different optimizers on the experimental results, the results are shown in Table 3.

Table 3 The impact of different optimizers on experimental results

| Optimizer | CTR | RR |
|---|---|---|
| AdaGrad | 0.76 | 0.80 |
| SGD | 0.74 | 0.80 |
| Momentum | 0.75 | 0.82 |
| Adam | 0.78 | 0.83 |

Table 3 shows the performance of different optimizers in the experiment, mainly using the two indicators of click-through rate and user retention rate to evaluate the performance of the model. It can be seen that different optimizers show different effects in the adaptive user interface generation task, which is closely related to the characteristics of the optimizer and its characteristics in parameter adjustment.

First, from the results of AdaGrad and SGD, the CTR is 0.76 and 0.74 respectively, and the RR is 0.80. AdaGrad is an optimizer suitable for sparse data. It can automatically adjust the learning rate, but it is easy to cause the learning rate to be too small when dealing with long-term training, which limits the further improvement of the model. Therefore, although it is slightly better than SGD in CTR, its overall performance is not outstanding. In contrast, SGD, as the most basic optimizer, has a simple update method and is usually prone to large fluctuations during training, affecting the convergence effect. These factors together lead to the fact that SGD and AdaGrad perform worse than other more complex optimizers in this task.

Secondly, the model using the Momentum optimizer achieved good results of 0.75 and 0.82 in CTR and RR respectively. Momentum effectively alleviates the oscillation problem of SGD by introducing the "momentum" of historical gradients during the update process, thereby accelerating the convergence speed. A higher RR indicates that the optimizer can better capture the user's long-term usage preferences and improve the stability and robustness of the model. However, in terms of CTR, the effect of Momentum is not optimal, indicating that the optimizer still has room for improvement in capturing the user's click preferences.

Third, the Adam optimizer performed best, with a CTR and RR of 0.78 and 0.83 respectively. This optimizer combines the advantages of momentum and adaptive learning rate adjustment, making it perform well in both convergence speed and effect. Adam can better adapt to the dynamic adjustment requirements of the user interface generation task, capture the user's subtle preferences, and also has a high user retention rate in long-term use. This shows that Adam can not only improve the attractiveness of the interface (CTR) in practical applications, but also significantly increase the user's stay time (RR), and is the most suitable optimizer choice for this task.

Therefore, in the adaptive user interface generation task, the Adam optimizer provides the best balance effect, which can bring users a more personalized and efficient interactive experience.

Finally, we present a line graph showing how resource utilization increases as the number of users increases, as shown in Figure 2.

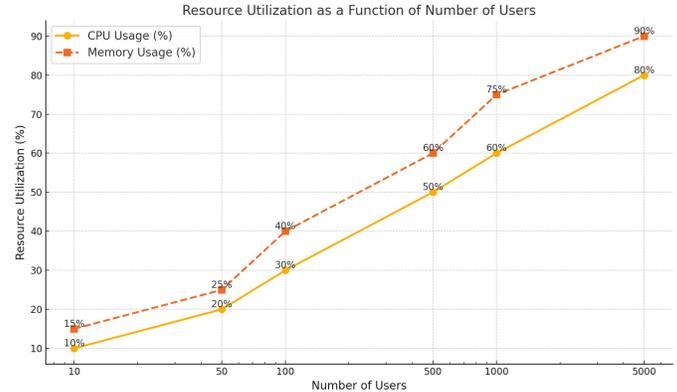

Figure 2 Resource Utilization Line Chart

As the number of users increases, the system's resource utilization gradually increases, reflecting the increasing demand for computing and memory in the reinforcement learning model in multi-user scenarios. The linear increase in CPU and memory utilization shows that the model can effectively scale to accommodate user loads of different sizes, while the higher memory usage reflects the complexity of the reinforcement learning model in storing interface states and processing user behaviors.

V. CONCLUSION

The adaptive user interface generation technology introduced in this study demonstrates significant effects and advantages in enhancing Human-Computer Interaction (HCI). By focusing on user behavior and interaction patterns, this approach aims to continuously learn and adapt to users' operating habits, enabling dynamic adjustment of interface layouts and configurations. The proposed system showed superior performance in terms of click-through rate and user retention rate compared to traditional methods such as multi-armed bandit and Bayesian optimization, highlighting its strength in delivering personalized and engaging user experiences.

This study further emphasizes the adaptability and scalability of adaptive interface generation systems in user interaction design. The system not only captures user behavior patterns effectively but also provides data-driven decision support for improving the quality and personalization of interfaces. Additionally, the integration of deep learning models, such as generative adversarial networks, can further enhance the personalization and quality of interface design. The future of adaptive user interface (UI) generation lies in its broader application across diverse scenarios, enabling more personalized and real-time human-computer interactions. Addressing key challenges, future research will focus on developing lightweight models and integrating edge computing to reduce resource demands while also improving system latency for real-time responsiveness [28]. Efforts will extend to adapting UIs for a wider range of devices and environments, fostering user trust through transparent data practices and enhanced privacy controls. Additionally, refining personalization techniques and optimizing system integration will be critical to delivering seamless and intuitive user experiences, establishing adaptive UIs as a cornerstone of advanced HCI.